\documentclass[11pt,a4paper]{article}

\usepackage{graphicx,epsfig,verbatim,enumerate}

\begin{document}

\title{The Synchronization of Clock Rate and the Equality of Durations Based on the
Poincar\'{e}-Einstein-Landau Conventions\footnote{Email:
Zhaoz43@hotmail.com}}
\author{Zhao Zheng$^1$ Tian Guihua$^2$   Liu
Liao$^1$ and Gao Sijie$^1$\\
1.Department of Physics,\\ Beijing Normal University, Beijing 100875, China\\
2.School of Science, Beijing University \\
of Posts And Telecommunications, Beijing 100876, China.}
\date{October 1, 2006}
\maketitle
\begin{abstract}

There are two important basic questions in the measurement of
time. The first one is how to define the simultaneity of two
events occurring at two different places. The second one is how to
define the equality of two durations. The first question has been
solved by Einstein, Landau and others on the convention that the
velocity of light is isotropic and it is a constant in empty
space. But no body has answered the second question until today.
In this paper, on the same convention about the velocity of light
given by Poincar\'{e}, Einstein, Landau and others, we find the
solution to the definition of the equality of two durations.
Meanwhile, we also find  answer to the question about the
definition of the synchronization of rate of clocks located at
different places.



\end{abstract}

\section{Introduction}

Both the simultaneity of clocks located at different places and the equality
of durations are important problems in research about time. Poincar\'{e}
thought that the two problems are related to each other, and that they can
only acquire meaning by convention. He guessed that one could solve the
problems by the convention that the velocity of light might be isotropic and
might be a constant in empty space\cite{1,2,3,4}\label{1,2,3,4}.

In the theory of the relativity, Einstein defined the simultaneity
of clocks located at different places by the convention that the
velocity of light is homogeneous and isotropic, and it is a
universal constant in empty space\cite {5,6}. Landau pointed out
that one can define globally the simultaneity of coordinate time
only in a time-orthogonal frame\cite{7}.

However they did not discuss how to define the equality of durations of a
clock.

We find that the synchronization of the moment of clocks located at
different places and the synchronization of the rate of these clocks are
different. The synchronization of the moment of clocks at different places
might be difficult, while synchronization of their rate turns out easier. We
suggested a new programme to synchronize the rate of clocks, where we do not
synchronize the moment of the clocks. Following the convention on the
velocity of light given by Poincar\'{e}, Einstein and Landau, we got the
condition of transitivity of synchronization of clock rate. In the
space-time where the condition is valid, we can define the identical ``clock
rate'' in the whole space-time, although there may not exist ``simultaneous
hypersurfaces''. In those papers, we stressed the synchronization of rates
of clocks located at different space points\cite{8,9,10,11}.

In this paper, with the programme to synchronize the rate of coordinate
clocks, we define the equality of durations of a clock by the above
convention on velocity of light. As a result, we not only answer the first
problem on the measurement of time advanced by Poincar\'{e} (i.e. the
definition on synchronization of clocks in different space points), but also
find the solution to his second problem (i.e. the definition of the equality
of durations).

The paper is organized as following. In the Sec.2, we introduce
Poincar\'{e}'s ideas about measurement of time. In the Sec.3 and
Sec.4, we present the results of research on simultaneity given by
Einstein, Landau and others. Then we give the result of research
on synchronization of rate of clocks in Sec.5. In Sec.6, we get
the definition on the equality of durations. In the Sec.7, short
conclusion and discussion are given.

\section{{Poincar\'{e}'s Notion of the Simultaneity and the Measurement of
Time}}

How to define the simultaneity of two events occurring at two different
places? How to define the equality of two durations? These two important
basic questions have been discussed by many famous philosophers, physicists
and mathematicians. Henri Poincar\'{e} gave inspired analysis on them before
the birth of A. Einstein's theory of relativity\cite{1,2,3,4}.

In ``The Measure of Time'' (1898) and ``The Value of Science'' (1905)
Poincar\'{e} pointed out: ``we have not a direct intuition of simultaneity
nor of the equality of two durations (time intervals)''. ``It is difficult
to separate the qualitative problem of simultaneity from the quantitative
problem of the measurement of time.''

``The qualitative problem of simultaneity is made of depend upon the
quantitative problem of the measurement of time.''

In ``Science and Hypothesis'' (1902) Poincar\'{e} wrote:

``(1). There is no absolute space, \ldots \ldots

(2)There is no absolute time. When we say that two periods are equal, the
statement has no meaning, and can only acquire a meaning by a convention.

(3)Not only have we no direct intuition of the equality of two periods, but
we have not even direct intuition of the simultaneity of two events
occurring at two different places.''

Poincar\'{e} thought that not only the equality of two durations but also
the simultaneity of two events occurring at two different places can only
acquire meaning by convention.

``This convention, however, is not absolutely arbitrary; it is not the child
of our caprice .We admit it because certain experiments have shown us that
it will be convenient, \ldots \ldots ''

``The simultaneity of two events, or the order of their succession, the
equality of two durations, are to be so defined that the enunciation of the
natural laws may be as simple as possible, \ldots \ldots ''

\section{The Definition on Simultaneity Given by Einstein}

A. Einstein agrees with Poincar\'{e}'s idea that the simultaneity of two
events occurring at two different places can only acquire meaning by
convention. Because ``physics'' is a science depending on experiment and
measurement, so any definitions based on convention about the simultaneity
must be operable in experiment and measurement. In the theory of
``relativity'', Einstein gave a definition of the simultaneity based on the
convention that the space is homogeneous and isotropic for light propagation
and that the velocity of light is a universal constant in empty space\cite
{4,5,6}.

In ``on the electrodynamics of moving bodies'' (1905) he wrote:

`We have so far defined only an ``A time'' and a ``B time''. We have not
defined a common ``time'' for A and B, for the latter cannot be defined at
all unless we establish by {\it definition} that the ``time'' required by
light to travel from A to B equals the ``time'' it requires to travel from B
to A. Let a ray of light start at the ``A time'' $t_A $ from A towards B.
Let it at the ``B time'' $t_B $ be reflected at B in the direction of A, and
arrive again at A at the ``A time''${t}^{\prime}_A $.

In accordance with definition the two clocks synchronize if

\begin{equation}
t_B-t_A=t_A^{\prime }-t_B  \label{1}
\end{equation}

We assume that this definition of synchronism is free from contradictions,
and possible for any number of points; and that the following relations are
universally valid:---

1. If the clock at B synchronizes with the clock at A, the clock at A
synchronizes with the clock at B.

2. If the clock at A synchronizes with the clock at B and also with the
clock at C, the clocks at B and C also synchronize with each other.

Thus with the help of certain imaginary physical experiments we have settled
what is to be understood by synchronous stationary clocks located at
different places, and have evidently obtained a definition of
``simultaneous,'' or ``synchronous,'' of ``time.'' The ``time'' of an event
is that which is given simultaneously with the event by a stationary clock
located at the place of the event, this clock being synchronous, and indeed
synchronous for all time determinations, with a specified stationary clock.

In agreement with experience we further assume the quantity
\begin{equation}
\frac{2AB}{t_A^{\prime }-t_A}=c  \label{2}
\end{equation}
to be a universal constant---the velocity of light in empty space.'

In Einstein's above theory, Eq.(\ref{1}) can be re-written as
\begin{equation}
\frac{\,t_A^{\prime }+t_A}2=t_B  \label{3}
\end{equation}
the ``A time'' $\widetilde{t_A}$%
\begin{equation}
\widetilde{t_A}=\frac{t_A^{\prime }+t_A}2  \label{4}
\end{equation}
is defined as the ``simultaneous'' moment with the ``B time '' $t_B$

With this way, Einstein defined ``simultaneity'' or ``synchronism'' of any
number of stationary clocks located at different places in a inertial frame.

It can be proved that the above definition and hypotheses are free
from contradictions in inertial frames.

However, the ``hypothesis 2'' may be invalid in an arbitrary reference
system in a curved space-time, or even in a non-inertial frame in a flat
space-time. It has been proved that the ``hypothesis 2'' is valid only in
time-orthogonal frames\cite{7}. It should be emphasized that even in a
time-orthogonal frame, in general, one can synchronize only the ``coordinate
clocks'' over all space, not the ``standard clocks''. In other words, one
only can globally define ``the simultaneity of the coordinate time'' but not
``the simultaneity of the proper time'' in time-orthogonal frames in general.

\section{The Condition of Transitivity of Simultaneity Given by Landau}

Let us discuss the possibility of synchronizing clocks at different points
over all space, i.e. the possibility of ``hypothesis 2'' being valid\cite
{7,8}.

Suppose a light signal travel from some point A in space to
another point B infinitely near to it and then back along the same
path. L and M are world lines of point A and point B respectively
(FiG. 1) . Let the light signal start at the ``A time'' $t_{A1}$
from A towards B, then it at the ``B time'' $t_B$ is reflected at
B in the direction of A and arrives again at A at the ``A time''
$t_{A2}{}_{.}$Let
\begin{equation}
dt_{(1)}=t_{A1}-t_B  \label{5}
\end{equation}
\begin{equation}
dt_{(2)}=t_{A2}-t_B  \label{6}
\end{equation}
\begin{equation}
t_A=\frac{t_{A1}+t_{A2}}2  \label{7}
\end{equation}

\begin{figure}[tbp]
\begin{center}
\includegraphics[clip,width=0.65\textwidth]{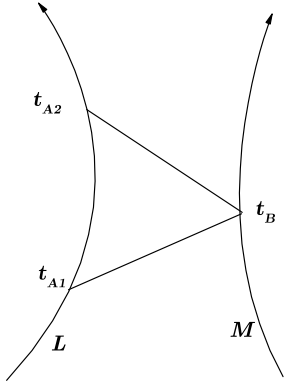}
\caption{The definition on the simultaneity of clocks located at
different places} \label{Fig.1}
\end{center}
\end{figure}

Following Einstein (Sec.2), one can define ``A time'' $t_A$ being the
simultaneous moment with ``B time''$t_B{}_{.}$ Substituting Eqs. (\ref{5})
and (\ref{6}) into Eq.(\ref{7}), one has
\begin{equation}
t_A=t_B+\frac 12\left[ dt_{(2)}+dt_{(1)}\right]   \label{8}
\end{equation}

Let us write the interval, separating the space and time coordinates, of two
events, as
\begin{equation}
ds^2=g_{00}dt^2+2g_{0i}dx^0dx^i+g_{ij}dx^idx^j,\quad (i,j=1,2,3)  \label{9}
\end{equation}

As the interval between the two events corresponds to the departure and
arrival of the light signal from one point to the other, it is equal to
zero. Solving the equation $ds^2=0$ with respect to $dx^0$, one finds two
roots:
\begin{equation}
dt=\frac{-g_{oi}dx^i\pm \sqrt{(g_{0i}g_{oj}-g_{00}g_{ij})dx^idx^j}}{g_{00}}
\label{10}
\end{equation}
corresponding to the propagation of the light signal in the two
directions between A and B. So, we have
\begin{equation}
dt_{(2)}+dt_{(1)}=-\frac{2g_{oi}}{g_{00}}dx^i{}^{.}  \label{11}
\end{equation}

Substituting Eq.(\ref{11}) into Eq.(\ref{8}), one gets
\begin{equation}
\Delta t\equiv t_A-t_B=-\frac{g_{0i}}{g_{00}}dx^i  \label{12}
\end{equation}

One finds that there exists a difference in the values of the coordinate
time t for two simultaneous moments at infinitely near points. Eq.(\ref{12})
enables people to synchronize clocks located at infinitely near points of
space. Carrying out a similar synchronization, one can synchronize clocks
located at different points alone any open curves of space, i.e. one can
define simultaneity of moments along any open curve. However synchronization
of clocks along a closed contour turns out to be impossible in general,
because $\triangle t$ is different from zero when one synchronizes clocks
starting out along the contour and returning to the initial point. In fact,
one has
\begin{equation}
\oint {\Delta t\ne 0}  \label{13}
\end{equation}

Hence it is impossible to synchronize clocks over all space unless the
reference system is time-orthogonal
\begin{equation}
g_{0i}=0  \label{14}
\end{equation}
or
\begin{equation}
\oint {\Delta t=\oint {\left( {-\frac{g_{0i}}{g_{00}}}\right) }}d{x^i=0}
\label{15}
\end{equation}
\section{Transitivity of Synchronization of Clock Rate}
In Ref.\cite{8,9,10,11}, we gave another kind of clock synchronization,
where we do not demand to synchronize ``simultaneous moments'' of coordinate
clocks, only demand to synchronize ``rates'' of coordinate clocks.

Now let us give the condition on the transitivity of synchronization of
clock rate. At the first simultaneous moment of the space points A and B,
the time difference of their coordinate clocks is
\begin{equation}
\Delta t_1=t_{A1}-t_{B1}=-\left( {g_{0i}/g_{00}}\right) _1dx^i,\quad
(i=1,2,3)  \label{16}
\end{equation}

At the second simultaneous moment, the time difference is
\begin{equation}
\Delta t_2=t_{A2}-t_{B2}=-\left( {g_{0i}/g_{00}}\right) _2dx^i,\quad
(i=1,2,3)  \label{17}
\end{equation}

The difference between the rates of the two clocks can be given:
\[
\delta \left( {\Delta t}\right) =\left( {\Delta t}\right) _A-\left( {\Delta t%
}\right) _B\equiv \left( {t_{A2}-t_{A1}}\right) -\left( {t_{B2}-t_{B1}}%
\right)
\]
\begin{equation}
=\left( {t_{A2}-t_{B2}}\right) -\left( {t_{A1}-t_{B1}}\right) =-\left[ {%
\left( {g_{0i}/g_{00}}\right) _2-\left( {g_{0i}/g_{00}}\right) _1}\right]
dx^i  \label{18}
\end{equation}

Therefore the rates of coordinate clocks are the same everywhere, or the
synchronization of clock rate is transitive, if and only if
\begin{equation}
\oint {\delta \left( {\Delta t}\right) =0}  \label{19}
\end{equation}
or
\begin{equation}
\oint {\left( {-\frac{g_{0i}}{g_{00}}}\right) _1}dx^i=\oint {\left( {-\frac{%
g_{0i}}{g_{00}}}\right) _2}dx^i  \label{20}
\end{equation}

That means
\begin{equation}
\frac \partial {\partial t}\oint {\Delta t=\frac \partial {\partial t}\oint {%
\left( {-\frac{g_{0i}}{g_{00}}}\right) }dx^i=0}  \label{21}
\end{equation}
or
\begin{equation}
\frac \partial {\partial t}\left( {-\frac{g_{0i}}{g_{00}}}\right) =0
\label{22}
\end{equation}

Eq.(\ref{21}) requires that the integral along a closed path is a
constant independent of time:
\begin{equation}
\oint {\Delta t=\oint {\left( {-\frac{g_{0i}}{g_{00}}}\right) }dx^i=}const
\label{23}
\end{equation}

Of course, the constant in the Eq.(\ref{23}) could be non-zero. This is a
weaker condition than that of time-orthogonality. Obviously, Eq.(21) is only
a necessary condition, but not a sufficient condition for constructing
simultaneity surfaces.

In the whole space--time which satisfies Eq. (\ref{23}) but not
Eq. (\ref {15}), one can synchronize all clocks such that rates of
the coordinate clocks are the same everywhere, or the
synchronization of clock rate is transitive. But the simultaneity
is not transitive, or one can not construct simultaneity surfaces
globally in the whole space-time.

\section{Measurement of Duration}

The simultaneity of two events occurring at two different places and the
transitivity of simultaneity have been defined, by Einstein, Poincar\'{e},
Landau and others on the foundation of the convention that the velocity of
light is homogeneous and isotropic, and is a universal constant in empty
space. Obversely, these important convention and definitions are operable in
experiment and measurement.

Now, let us discuss the issue about ``the equality of two durations'', i.e.
``the equality of two periods or two time intervals''. Poincar\'{e} thought
that this problem of the measurement of time is related to the problem of
simultaneity.

We want to point out that we can define ``the equality of two
durations'' based on the same convention about the velocity of
light in empty space, and that this definition is also operable in
experiment and measurement.

\subsection{Stationary Space-times}

Let an observer with a light source and a mirror rest at a fixed
space point A in a stationary space-time, and another observer
with a mirror rest at the point B infinitely near to A. L and M
are respectively the world-lines of the observers resting at the
points A and B (FIG.2). Let the observer A emit a light signal at
the ''A time'' $t_{A1}$ towards B. The signal is reflected by the
mirror of the observer B at the ``B time'' $t_{B1}$ in the
direction of A, and the light signal arrives again at A at the ``A
time '' $t_{A2}$. When the light signal goes to B and back to A,
the difference of the coordinate time is
\begin{equation}
\Delta t_1=t_{A2}-t_{A1}  \label{24}
\end{equation}

\begin{figure}[tbp]
\begin{center}
\includegraphics[clip,width=0.65\textwidth]{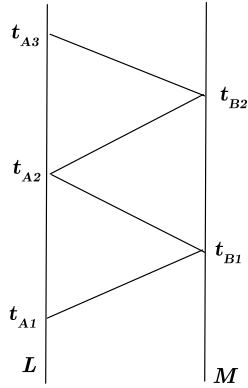}
\caption{The definition on ''the equality of durations'' in a
stationary space time} \label{Fig.2}
\end{center}
\end{figure}
At the same moment ($t_{A2})$ receiving the light signal reflected
by B, the mirror A reflects again the signal in the direction of B.
The light signal arrives again at B at the ``B time''$t_{B2}$, then
it is reflected and comes back to A at the ``A time''$t_{A3}$,\ldots
\ldots . The travel of the light signal to and fro between A and B
can be regarded as a period of time. The duration of the first
period is shown in Eq.(\ref{24}). The duration of second period is:
\begin{equation}
\Delta t_2=t_{A3}-t_{A2}  \label{25}
\end{equation}
\[
\vdots
\]
The duration of the nth period is:
\begin{equation}
\Delta t_n=t_{A(n+1)}-t_{An}  \label{26}
\end{equation}
There is no any reason to think that the ``durations'' of the above periods
are different, because the space-time is stationary. So, one has

\begin{equation}
\Delta t_1=\Delta t_2=\cdot \cdot \cdot =\Delta t_n  \label{27}
\end{equation}
i.e.
\begin{equation}
t_{A2}-t_{A1}=t_{A3}-t_{A2}=\cdot \cdot \cdot =t_{A(n+1)}-t_{An}  \label{28}
\end{equation}

Consequently, we give a definition on the equality of durations in a
stationary space-time.

Discussion:

(i) Because the distance between the point B and the point A is arbitrary
selected, we can define the equality of durations of any length.

(ii) The durations discussed above are the durations of
coordination time. The relation between the duration of coordinate
time dt and the duration of proper time d$\tau $ is
\begin{equation}
d\tau =\sqrt{-g_{00}}dt  \label{29}
\end{equation}

We have the conclusion that the durations of the proper time of a
resting observer, shown by Eq.(\ref{29}), are equal, that is
\begin{equation}
d\tau _1=d\tau _2=\cdot \cdot \cdot =d\tau _n,  \label{30}
\end{equation}
because $g_{00}$ is independent of time t in stationary space times.

(iii) We can generalize the definition on ``the equality of
durations'' to the whole space-time, and get the conclusion that
the durations of coordinate time are equal in the whole stationary
space--time, because the condition ((\ref{19}) or (\ref{20})) of
``transitivity of synchronization of clock rate'' is valid in any
stationary space-times. From Eq. (\ref{29}), we know that the
durations of proper time of the different observers located at the
different space points are not equal in general, owing to $g_{00}$
being a function depending on space points.

\subsection{A General Case}

Let $\mathcal L$ be a congruence of the world-lines of a family of
static observers A, B, C, \ldots in a curved space-time (FIG.3).
It means that a coordinate system covering $\mathcal L$ is
selected, where $\mathcal L$ is just the congruence of the time
coordinate curves. We assume the synchronization of clock rates is
transitive, but the coordinate system is not time-orthogonal.

Let us synchronize clocks along a closed space contour with a light signal
following Einstein and Landau. We know from Eq. (\ref{12}) that the ``B
time'' $t_B$ of the point B near A is defined as the simultaneous moment of
the ``A time'' $t_{A1}$ when the time difference of the two simultaneous
moments is given as
\begin{equation}
\Delta t=-\frac{g_{0i}}{g_{00}}dx^i  \label{31}
\end{equation}

\begin{figure}[tbp]
\begin{center}
\includegraphics[clip,width=0.65\textwidth]{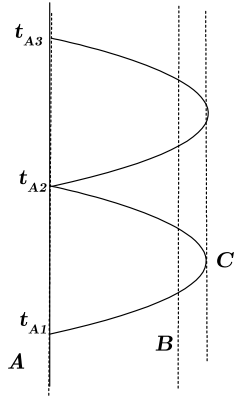}
\caption{The definition on ''the equality of durations'' in a
general spacetime}
 \label{Fig.3}
 \end{center}
\end{figure}

Similarly, we can define the ``C time'' $t_C$ of the point C near B as the
simultaneous moment of the ``B time'' $t_B$,\ldots \ldots . When we
synchronize clocks along the closed space contour returning to the initial
point A, we find that the simultaneous moment ,shown by the clock A, is $%
t_{A2}$, which is different from $t_{A1}$, the difference of the two
simultaneous moments is
\begin{equation}
\Delta t_1=t_{A2}-t_{A1}=\oint_1{\left( {-\frac{g_{0i}}{g_{00}}}\right) dx^i}%
=a_1  \label{32}
\end{equation}

Using the same method, we continue to synchronize clocks along the same
closed contour, we have
\begin{equation}
\Delta t_2=t_{A3}-t_{A2}=\oint_2{\left( {-\frac{g_{0i}}{g_{00}}}\right) dx^i}%
=a_2  \label{33}
\end{equation}
\[
\vdots
\]
\begin{equation}
\Delta t_n\equiv t_{A(n+1)}-t_{An}=\oint_n{\left( {-\frac{g_{0i}}{g_{00}}}%
\right) dx^i}=a_n  \label{34}
\end{equation}

Owing to the coordinate system being not time --orthogonal, we have
\begin{equation}
a_i\ne 0,\quad (i=1,2,\cdots ,n)  \label{35}
\end{equation}
But we know that, from Eqs.(\ref{19})-(\ref{23}),
\begin{equation}
a_1=a_2=\cdots =a_n=a=const  \label{36}
\end{equation}

Because the synchronization of clock rate is transitive in the coordinate
systems, we get the equal time periods from $t_{A1} $ to $t_{A2} $, from $%
t_{A2} $ to $t_{A3} $, etc. We can define the equal periods as the
equal durations. Thus we obtain the equal ''time intervals'' of
the observer A.

It is easy to know that the constant $a$ in Eq.(\ref{36}) will be
different when we synchronize clocks along different closed space
contours. Because we can select arbitrary contour to synchronize
clocks, so we can get arbitrary value of $a_{.}$It means that we
can define arbitrary length of ''equality of duration'', i.e.
arbitrary length of ''equality of time intervals''. Certainly,
that is the duration of coordinate time.

The relation between the coordinate time t and the proper time $\tau $ of
the observer A is
\begin{equation}
d\tau =\sqrt{-g_{00}}dt  \label{37}
\end{equation}
because A rests in the coordinate system. Therefore, we can define
''length'' of the durations of proper time by means of
\begin{equation}
\Delta \tau_k=\int_{ka}^{(k+1)a}\sqrt{-g_{00}}dt.  \label{38}
\end{equation}

In other words, we can parameterize the world line with the duration of proper time $%
\Delta \tau_k $ given by Eq.(\ref{38}).

\section{Conclusion and discussion}

It should be noticed that Eq.(36) is based upon the condition on the
transitivity of synchronization of clock rate. Our original purpose
suggesting the notion of ``the transitivity of synchronization of clock
rate'' is to find the coordinate clocks whose rates are the same in the
whole space. It means that the ``time periods'' of the same kind of periodic
motion are the same through the whole space time. Here we want to emphasize
that the condition on the transitivity of synchronization of clock rate is
not only the condition of the equality of the rates of clocks located at
different space points, but also the condition of the equality of durations
of a clock. Therefore, the condition not only solves the question about the
definition of the synchronization of rate of clocks located at different
space points, but also solves the question about the definition of the
equality of two durations. It means that the condition on transitivity of
synchronization of clock rate is the condition to define identical, equal
time intervals through the whole space time.

\section*{Acknowledgments}

This research is supported by the National Natural Science
Foundations of China (grant No.10373003 {\&} No.10475013) the
National Key Basic Research and Development Programme of China (
grant No. 2003CB716300). We would like to thank Prof. Zhu
Jian-yang, Mr. Liu Jie-Min, Yuan Ping, Zhao Xue-Song, Chen Ping
and Zhong Shu-Quan for Their help.


\begin{thebibliography}{99}
\bibitem{1} H. Poincar\'{e}, Science and Hypothesis, London{: }Walter
Scott Publishing (1905) 89-110, 123-139
\bibitem{2} H. Poincar\'{e}, The Value of science, New York: Dover , 1958,
223-234.
\bibitem{3} H Poincar\'{e}, The Foundation of Science, New York: The
Science Press, 1913.
\bibitem{4}  A I Miller Albert Einstein's Special Theory of Relativity,
London: Addison-Wesley Publishing Company, Inc. 1981, 185-200.
\bibitem{5}  A Einstein, The Principle of Relativity, Dover: Dover
Publications, 1923.
\bibitem{6}  A Einstein, The Meaning of Relativity, Princeton: Princeton
University Press, 1950.
\bibitem{7}  L D Landau, E M Lifshitz, The classical Theory of Fields,
Beijing: World Pubbishing Corporation, 1999.
\bibitem{8}  . Liu Liao, Zhao Zheng, General Relativity (2nd ed.), Beijing:
Higher education Press, 2004, 21-23 (in Chinese).
\bibitem{9}  Zhao Zheng, The transitivity of thermal equilibrium and the
transitivity of the clock rate synchronization, Science in China
{\bf A34} (1991) 835-840.
\bibitem{10}  Zhao Zheng, Ping Chen, Zeroth law of
thermodynamics and transitivity of simultaneity, Int. J. Theo.
Phys. {\bf 36 }(1997) 2153.
\bibitem{11}  Zhao Zheng, Pei Shou-Yong and Liu Liao, Transitivity of the clock rate
synchronization being equivalent to the zeroth law of thermodynamics, Acta.
Physica. Sinica {\bf 48} (1999) No.11. 2004-2010 (in Chinese).
\end{thebibliography}
\end{document}